\documentclass[twocolumn,showpacs,preprintnumbers,amsmath,amssymb,A4paper]{revtex4}

\usepackage{graphicx}
\usepackage{dcolumn}
\usepackage{bm}

\begin{document}
\newcommand{\kp}{{\bf k$\cdot$p}\ }
\newcommand{\Pp}{{\bf P$\cdot$p}\ }

\preprint{APS/123-QED}

\title{Spin and cyclotron energies of electrons in GaAs/Ga$_{1-x}$Al$_x$As quantum wells}

\author{P. Pfeffer and W. Zawadzki}
 \affiliation{Institute of Physics, Polish Academy of Sciences,\\
Al.Lotnikow 32/46, 02-668 Warsaw, Poland\footnotetext{Corresponding author: P.
Pfeffer, pfeff@ifpan.edu.pl}}

\date{\today}

\begin{abstract}
A five-level {\Pp} model of the band structure for GaAs-type semiconductors is used to describe the spin
$g^*$-factor and the cyclotron mass $m^*_c$ of conduction electrons in GaAs/Ga$_{1-x}$Al$_x$As quantum wells in
an external magnetic field parallel to the growth direction. It is demonstrated that the previous theory of the
$g^*$-factor in heterostructures is inadequate. Our approach is based on an iteration procedure of solving 14
coupled differential {\Pp} equations. The applicability of the iteration procedure is verified. The final
eigenenergy problem for the conduction subbands is reduced to two differential equations for the spin-up and
spin-down states of consecutive Landau levels. It is shown that the bulk inversion asymmetry of III-V compounds
is of importance for the spin $g^*$-factor. Our theory with no adjustable parameters gives an excellent
description of experimental data on the electron spin $g^*$-factor in GaAs/Ga$_{0.67}$Al$_{0.33}$As rectangular
quantum wells for different well widths between 3 and 12 nm. The same theory describes very well experimental
cyclotron masses in GaAs/Ga$_{0.74}$Al$_{0.26}$As quantum wells for the well widths between 6 and 37 nm.
\end{abstract}

\maketitle

\section{\label{sec:level2} INTRODUCTION \protect\\{}}

Spin properties of electrons in semiconductor heterostructures have become in recent years subject of intense
experimental and theoretical interest because of their inherent scientific value as well as possible spintronic
applications. Among numerous heterostructures, the system GaAs/Ga$_{1-x}$Al$_x$As has won a unique position.
First, GaAs is after silicon the most important semiconductor material. Second, the system
GaAs/Ga$_{1-x}$Al$_x$As has well established parameters which allows the theorists to describe new subtle
phenomena. Third, due to the advanced growth technology the electrons in GaAs quantum wells have very high
mobilities making it possible to detect even weak effects. Thus it is of interest to describe precisely the spin
and orbital electron energies in GaAs/Ga$_{1-x}$Al$_x$As quantum wells not only for their own sake, but also as
a model example for other heterostructures. From the theoretical point of view, a description of magnetooptical
effects in GaAs-type materials is not an easy task since GaAs is a medium gap semiconductor. As a result, its
band structure exhibits features characteristic of narrow-gap materials, but it is insufficient to treat them by
the models normally used for such systems. We showed in our previous work on bulk GaAs and InP that the simplest
adequate description of the band structure of these materials is given by the so called five-level {\Pp} model
[1, 2]. Thus, the present work on the spin $g^*$-factor of electrons in GaAs/Ga$_{1-x}$Al$_x$As quantum wells
was first motivated by the fact that the theoretical description used in the literature for this purpose had
been based on the three-level model [3]. Examining the problem we realized that the theory [3] suffers from
other deficiencies, as we will demonstrate below. The theory we develop for the spin energies is equally valid
for the cyclotron (orbital) energies, so that we describe also the cyclotron masses, although here our treatment
improves only slightly the existing approach [4].

An important feature of the III-V compounds is a bulk inversion asymmetry (BIA) of these materials. As is known
from the fifties [5], the BIA results in a spin splitting of energies for a given direction of the wavevector
\textbf{k}. It is, however, of interest to investigate how this spin splitting behaves in the presence of an
external magnetic field. One can rephrase the problem by asking how the Zeeman spin splitting caused by the
magnetic field combines with the spin splitting caused by BIA. In the bulk magneto-optical or magneto-transport
studies one is usually interested in the Landau levels at $k_z$ = 0, since the density of Landau states has
singularities at the vanishing $k_z$. However, in a quantum well one deals with electric subbands for which the
value of $k^2_z$ is "frozen" in the subband wavefunction along the growth direction. The BIA splitting is
sensitive to this $k^2_z$ value which increases with a decreasing well width. This in turn is reflected in the
spin $g^*$-factor. In the following we will be concerned with symmetric quantum wells, so that the
Bychkov-Rashba spin splitting, caused by the structure inversion asymmetry, does not come into play [6].

Our calculation of the spin and cyclotron electron energies in GaAs/Ga$_{1-x}$Al$_x$As heterostructures is based
on the five-level {\Pp} model (5LM). An important advantage of the model is that it includes the BIA mechanism
of the spin splitting. We solve the eigenenergy problem by an iteration procedure which is more precise than an
expansion in powers of momentum used in the literature. We check the precision of consecutive iteration steps.
The results are compared with those obtained by the three-level {\Pp} model (3LM) and it is shown that the
latter is insufficient for the GaAs-type materials.

Our paper is organized as follows. In Section II the {\Pp} theory within 5LM is formulated for the bulk
semiconductor. Next the 3LM model is used to obtain results for the spin $g^*$-value. It is demonstrated that
the procedure of Ref. [3] based on the same model is inadequate. The iteration solution of 5LM is checked
against "exact" results for the bulk, as obtained by a numerical procedure. In Section III the theory for the
quantum wells is worked out from the 5LM matrix. Results for the spin $g^*$-factor and the cyclotron mass are
presented, compared with experimental data and discussed in Section IV. The paper is concluded by a summary.

\section{\label{sec:level2}P$\cdot$\lowercase{p} THEORY\protect\\{}}

 In order to discuss the previous treatments and to establish standards of
precision for various approximations we first consider the 3D case. The {\Pp} theory, which is the {\kp} theory
generalized for the presence of an external magnetic field \textbf{B}, has the form (cf. Ref. [1])
$$
\sum_l\left[\left( \frac{P^2}{2m_0}+E^{(l)}-{\cal E}\right)\delta_{l'l}+\frac{1}{m_0}{\bf p}_{l'l}\cdot\bf{P}+
\right.
$$
\begin{equation}
\left. +\mu_{\rm B} {\bf B} \cdot{\bm \sigma}_{l'l}+H^{so}_{l'l} \right]f_l =0
\end{equation}
where ${\cal E}$ is the energy, ${\bf P} = {\bf p}+e{\bf A}$ is the kinetic momentum, ${\bf A}$ is the vector
potential of magnetic field $\bf B$, and ${\bm \sigma}_{l'l} = (1/\Omega)<u_{l'}|{\bm \sigma}|u_l>$. Here ${\bm
\sigma}$ are the Pauli spin matrices, $\Omega$ is the volume of the unit cell, $u_l$ are periodic amplitudes of
the Luttinger-Kohn functions, $\mu_B = e\hbar/2m_0$ is the Bohr magneton, ${\bf p}_{l'l}$ are the interband
matrix elements of momentum and $H^{so}_{l'l}$ are those of spin-orbit interaction. The sum in Eq. (1) runs over
all bands $l = 1, 2,...$ included in the model, $l'=1, 2,...$ runs over the same bands, $E^{(l)}$ are the
band-edge energies, see below.

\begin{figure}
\includegraphics[scale=0.55,angle=0]{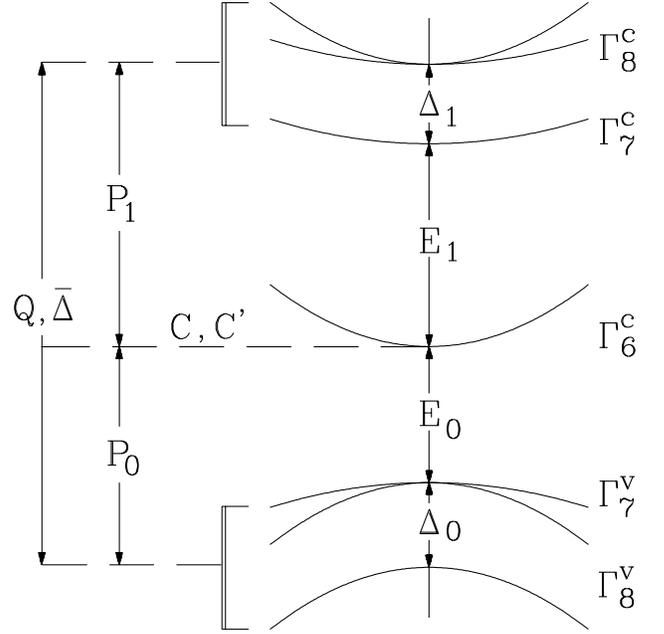}
\caption{\label{fig:epsart}{Five-level model for the band structure of GaAs-type semiconductors. Energy gaps,
spin-orbit energies, interband matrix elements of momentum and of the spin-orbit interaction are indicated.
Letters C and C' mark symbolically far-band contributions to the effective mass and the spin $g^*$-factor of
conduction electrons, respectively.}} \label{fig1th}
\end{figure}

 In the following we shall be concerned with the five-level model of the band
structure, as illustrated in Fig. 1. Within this model there exist three interband matrix elements of momentum
\begin{equation}
 P_0=\frac{-i\hbar}{m_0\Omega}<S|p_x|X> \;\;\;,
 \end{equation}
\begin{equation}
 P_1=\frac{-i\hbar}{m_0\Omega}<S|p_x|X'> \;\;\;,
 \end{equation}\begin{equation}
 Q=\frac{-i\hbar}{m_0\Omega}<X|p_y|Z'>=\frac{i\hbar}{m_0\Omega}<X'|p_y|Z> \;\;\;,
 \end{equation}
and three matrix elements of the spin-orbit interaction
\begin{equation}
 \Delta_0=\frac{-3i\hbar}{4m_0^2 c^2}<X|[{\bm \nabla} V_0,{\bf p}]_y|Z> \;\;\;,
\end{equation}

\begin{equation}
 \Delta_1=\frac{-3i\hbar}{4m_0^2 c^2}<X'|[{\bm \nabla} V_0,{\bf p}]_y|Z'> \;\;\;,
\end{equation}

\begin{equation}
 \overline{\Delta}=\frac{-3i\hbar}{4m_0^2 c^2}<X|[{\bm \nabla} V_0,{\bf p}]_y|Z'> \;\;\;,
\end{equation}
where the nonprimed functions are related to the $\Gamma^v_7$, $\Gamma^v_8$ valence bands, while the primed
functions are related to the $\Gamma^c_7$, $\Gamma^c_8$ conduction bands. The band edge energies $E^{(l)}$ are
generally influenced by $\overline {\Delta}$ (cf. Ref. 1) but in
\begin{figure}
\includegraphics[width=0.99\textheight, angle=90]{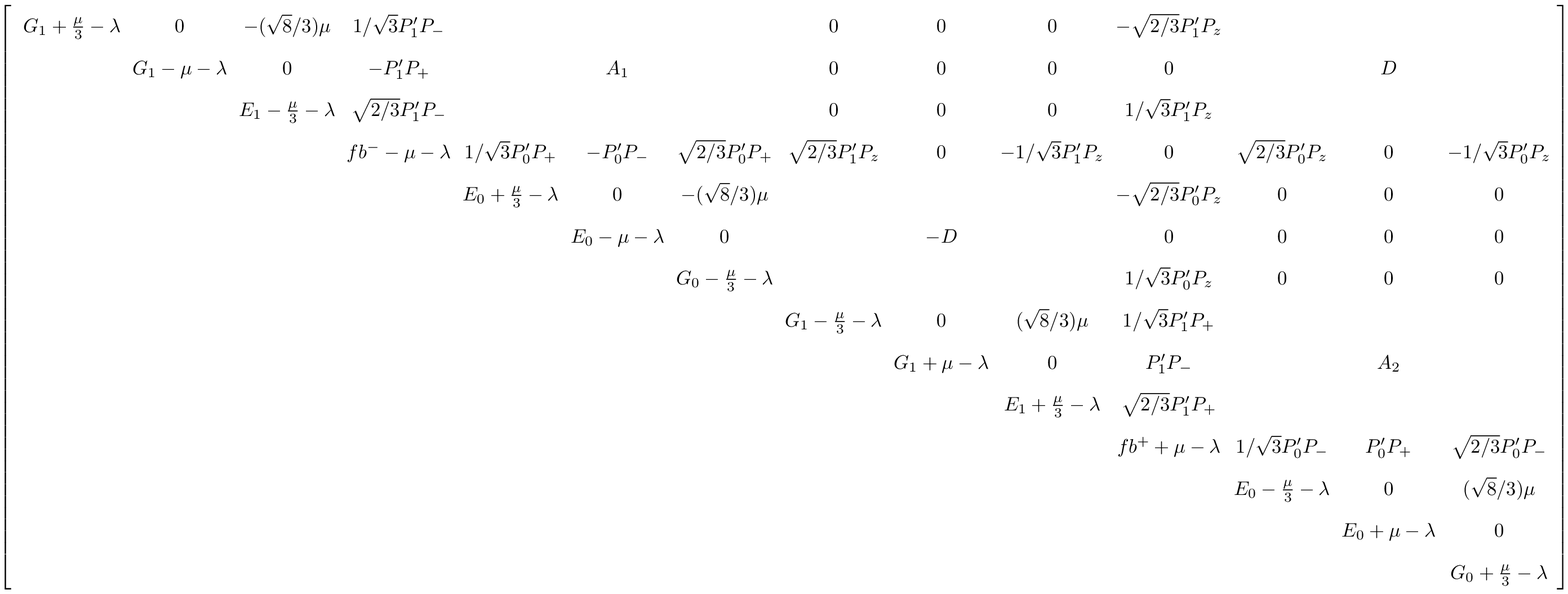}
\hfill(8)
\end{figure}
\stepcounter{equation} GaAs-type materials, where $\overline {\Delta}$ is small, this feature may be ignored and
we have: $E^{(1)} = E^{(2)} = E^{(8)} = E^{(9)} = G_1 = E_1+\Delta_1;\; E^{(3)} = E^{(10)} = E_1;\; E^{(5)} =
E^{(6)} = E^{(12)} = E^{(13)} = E_0;\; E^{(7)} = E^{(14)} = G_0 = E_0+\Delta_0$. We take the zero of energy at
the $\Gamma^c_6$ conduction band edge. Within 5LM there are 14 basis functions and the {\Pp} differential
Hamiltonian for the envelope functions $f_l(\bf{r})$ has the explicit form of Eq. (8). We define $P_0^{'} =
P_0/\hbar$, $P_1^{'} = P_1/\hbar$, $Q^{'} = Q/\hbar$, $\lambda = {\cal
E}-\hbar\omega_c^0(a^{+}a+\frac{1}{2})-p_z^2/2m_0$. Here $\omega_c^0 = eB/m_0$ is the cyclotron frequency with
free electron mass. The quantities $fb^{\pm} = C[\hbar\omega_c^0(a^{+}a+\frac{1}{2})+p_z^2/2m_0]\pm C'\mu_BB$
result from the far band contributions to the conduction band energies. The operators $P_{\pm} = (P_{x}\pm
iP_{y})/\sqrt{2}$ are proportional to the raising and lowering operators for the harmonic oscillator functions:
$P_{+} = -(\hbar/L)a^{+}$ and $P_{-} = -(\hbar/L)a$, in which $L = (\hbar/eB)^{1/2}$ is the magnetic radius. The
matrices $A_1$, $A_2$ and $D$ have different forms for various directions of magnetic field. In the following we
shall be concerned with the case \textbf{B}$\||$[001] crystal direction, for which
\begin{equation}
A_1=A_2= \left[ \begin{array}{ccc}
\frac{1}{3}\overline{\Delta}&\frac{1}{\sqrt 3}Q'P_z&0 \\
-\frac{1}{\sqrt 3}Q'P_z&\frac{1}{3}\overline{\Delta}&-\sqrt{\frac{2}{3}}Q'P_z \\
0&\sqrt{\frac{2}{3}}Q'P_z&-\frac{2}{3}\overline{\Delta} \\
\end {array}
 \right]\;\;,
\end{equation}
\ \\
\begin{equation}
D= \left[ \begin{array}{ccc}
0&\-\sqrt{\frac{2}{3}}Q'P_-&Q'P_+ \\
-\sqrt{\frac{2}{3}}Q'P_-&0&\sqrt{\frac{1}{3}}Q'P_- \\
-Q'P_+&\sqrt{\frac{1}{3}}Q'P_-&0 \\
\end {array}
 \right]\;\;.
\end{equation}

For $Q$=0 and $k_z$=0 matrix (8) factorizes into two 7$\times$7 matrices containing spin-up and spin-down
electron energies. For $Q$ = 0 and $k_z \ne$ 0 the resulting bands are spherical (i.e. the energies do not
depend on the direction of \textbf{B}), but the spin-up and spin-down states are mixed by the $k_z$ terms.
Still, in this case one can find solutions of the eigenvalue problem in form of single harmonic oscillator
functions. If $Q \ne$0, the resulting bands are nonspherical and the above simple solutions do not exist. At
this point we emphasize that matrix (8) contains no approximations within 5LM.

For considerations of the bulk there is no external potential. To describe the magnetic field we take the gauge
\textbf{A}=[-By, 0, 0], corresponding to \textbf{B}$||$[001]. One can then look for the envelope functions in
the form $f_l = exp(ik_xx+ik_zz)\Phi_l(y)$, where $\Phi_l(y)$ are the harmonic oscillator functions.

\subsection{Three-level model}

The three level model of $\Gamma^c_6$, $\Gamma^v_7$, $\Gamma^v_8$ levels (see Fig. 1) is not adequate for the
description of the conduction band in GaAs - type materials (see [1, 2]), but it can be solved exactly and we
use it as a starting point of our considerations as well as for a discussion of the existing work on the
$g^*$-values.

Once the higher conduction levels are omitted, the couplings $P_1$, $Q$ and $\overline{\Delta}$ do not come into
play (see Fig. 1), the conduction band is spherical and the resulting 8$\times$8 Hamiltonian may be solved in
terms of eight harmonic oscillator functions. This was first done by Bowers and Yafet [7], see also [8, 9]. In
our more general formulation (8), the 3LM corresponds to considering the columns and rows 4, 5, 6, 7, 11, 12,
13, 14. The resulting energy for the spin-up state of the n-th Landau level is given by

$$ {\cal
E}=\hbar\omega^0_c(n+\frac{1}{2})\left[1+C-\frac{E_{P_0}}{3}\left(\frac{3}{2E^0_1}+
\frac{1}{2E^0_3}+\frac{1}{G^0_2}\right)\right]+
$$
$$
+\frac{1}{2}\mu_BB\left[2+2C'+\frac{2E_{P_0}}{3}\left(\frac{3}{2E^0_1}-\frac{1}{2E^0_3}-
\frac{1}{G^0_2}\right)\right]+
$$
\begin{equation}
+\frac{\hbar^2k^2_z}{2m_0}\left[1+C-\frac{E_{P_0}}{3}\left(\frac{2}{E^0_2}+\frac{1}{G^0_1}\right)\right]\;\;,
\end{equation}
while the energy for the spin-down state of the n-th Landau level is given by
$$
{\cal E}=\hbar\omega^0_c(n+\frac{1}{2})\left[1+C-\frac{E_{P_0}}{3}\left(\frac{3}{2E^0_4}+
\frac{1}{2E^0_2}+\frac{1}{G^0_1}\right)\right]+
$$
$$
-\frac{1}{2}\mu_BB\left[2+2C'+\frac{2E_{P_0}}{3}\left(\frac{3}{2E^0_4}-\frac{1}{2E^0_2}-
\frac{1}{G^0_1}\right)\right]+
$$
\begin{equation}
+\frac{\hbar^2k^2_z}{2m_0}\left[1+C-\frac{E_{P_0}}{3}\left(\frac{2}{E^0_3}+\frac{1}{G^0_2}\right)\right]\;\;,
\end{equation}
where
$$
E^0_1=E_0+\mu_BB+\hbar\omega^0_c(n-\frac{1}{2})-\cal{E}\;\;,
$$
$$
E^0_2=E_0+\frac{1}{3}\mu_BB+\hbar\omega^0_c(n-\frac{1}{2})-\cal{E}\;\;,
$$
$$
E^0_3=E_0-\frac{1}{3}\mu_BB+\hbar\omega^0_c(n+\frac{3}{2})-\cal{E}\;\;,
$$
$$
E^0_4=E_0-\mu_BB+\hbar\omega^0_c(n+\frac{3}{2})-\cal{E}\;\;,
$$
$$
G^0_1=G_0-\frac{1}{3}\mu_BB+\hbar\omega^0_c(n-\frac{1}{2})-\cal{E}\;\;,
$$
\begin{equation}
G^0_2=G_0+\frac{1}{3}\mu_BB+\hbar\omega^0_c(n+\frac{3}{2})-\cal{E}\;\;.
\end{equation}
The first terms in Eqs. (11) and (12) represent the orbital parts, the second are related to the spin parts
(they differ in sign for spin-up and spin-down states) and the third give $k_z^2$ parts for the motion along the
magnetic field. Equations (11) and (12) represent fourth-order polynomials for the unknown energies. If we
neglect the free-electron and far-band contributions, Eqs. (11) and (12) reduce to cubic equations for the
energies. An influence of free electron terms on the description of spin $g^*$-factors in selected materials was
discussed by Singh et al [10].

In investigations of bulk semiconductors one is usually interested in the simplified case of $k_z$ = 0 since the
singular density of states for the Landau levels corresponds to the vanishing $k_z$. However, our final aim in
this work is to investigate quantum wells, which corresponds to the situation of $k_z \ne$ 0. Solving Eqs. (11)
and (12) for the spin-up and spin-down energies one can find the cyclotron effective mass and the spin
$g^*$-value exactly within 3LM. We define the cyclotron mass $m^*_c$ and the spin $g^*$-value in the standard
way
\begin{equation}
{\cal E}_{n+1}^{\pm} - {\cal E}_n^{\pm}=\frac{\hbar eB}{m^*_c}\;\;,
\end{equation}

\begin{equation}
{\cal E}_n^+ - {\cal E}_n^-=\mu_Bg^*B\;\;.
\end{equation}

The dependence of $m^*_c$ and $g^*$ on the electron energy was established in the early days, see Refs [11, 12].
Knowing the solutions of Eqs. (11), (12) and fixing magnetic field intensity $B$ one can determine the
dependence of $m^*_c$ and $g^*$ on $k_z^2$ exactly within 3LM.

Since the conduction band in GaAs is only weakly nonparabolic, we can expand the exact energies using Eqs. (11)
and (12) in the limit ${\cal E}/E_0 << 1$. The zero-order terms give $m^*_0$ and $g^*_0$ values at the
conduction band edge. The first-order terms give the first nonparabolic approximation. One can alternatively say
that the zero-order terms are proportional to $P_0^2$ while the first-order terms are proportional to $P_0^4$.
Using Eqs. (11), (12) and the definitions (14) and (15) we obtain after some manipulation (we keep the free
electron terms and the far-band contributions only in the conduction band)
\begin{equation}
\frac{1}{m^*_c}=\frac{1}{m^*_0}\left[1-\frac{E_{P_0}}{3}D_{nk_z}\left(\frac{1}{G_0^2}+\frac{2}{E_0^2}\right)\right]\;\;,
\end{equation}

$$
g^*=g^*_0+D_{nk_z}\frac{E_{P_0}}{3}\left[\frac{2m_0}{m^*_0}\left(\frac{1}{E_0^2} -\frac{1}{G_0^2}\right)+
\right.
$$
\begin{equation}
\left. -g^*_0\left(\frac{1}{G_0^2}+\frac{2}{E_0^2}\right)\right]\;\;,
\end{equation}
where $D_{nk_z}=\hbar\omega_c^0(n+1/2)+\hbar^2k_z^2/2m_0$, and
\begin{equation}
\frac{m_0}{m^*_0}=1+C-\frac{E_{P_0}}{3}\left(\frac{2}{E_0}+\frac{1}{G_0}\right)\;\;,
\end{equation}
\begin{equation}
g^*_0=2+2C'+\frac{2E_{P_0}}{3}\left(\frac{1}{E_0}-\frac{1}{G_0}\right)\;\;,
\end{equation}
are the band-edge values of the effective mass and of the $g^*$-factor according to 3LM. We define $E_{P_0} =
2m_0P^2_0/\hbar^2$. To calculate the numerical values we take the following band parameters for GaAs: $E_{P_0}$
= 27.865eV, $E_0$ = -1.519 eV, $G_0$ = -1.86 eV, $C$ = -3.070, $C'$ = -0.102. This results in $m^*_0$ = 0.0660
$m_0$ [2] and $g^*_0$ = -0.44 [13].

We would like to plot the $g^*$ value as a function of energy in view of the applications to quantum wells (QW).
To this end we calculate the theoretical energy $\cal E$ for the ground electric subband in a rectangular
Ga$_{0.67}$Al$_{0.33}$As/GaAs/Ga$_{0.67}$Al$_{0.33}$As QW as a function of the well width $d$. Next we define
the value of $k_{z0}$ by the equality: ${\cal E}(d) = \hbar^2k_{z0}^2/2m^*_0$, where $m^*_0$ = 0.0660 $m_0$.
This value of $k_{z0} = (2m^*_0{\cal{E}}/\hbar^2)^{1/2}$ is introduced to the above relations for $g^*$. When
describing the final results for the $g^*$-value and the cyclotron mass in QW we will use the same range of well
widths.

\begin{figure}
\includegraphics[scale=0.55,angle=0]{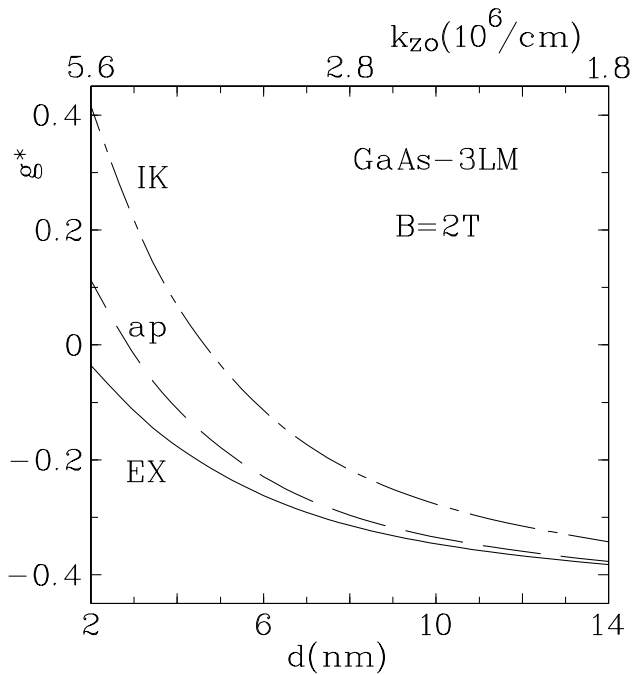}
\caption{\label{fig:epsart}{Theoretical spin $g^*$-factor of conduction electrons in bulk GaAs \emph{versus} the
wave vector $k_{z0}$ parallel to the magnetic field \textbf{B}, or the corresponding well width $d$ (see text).
The solid line (EX) shows exact results from the three-level model, the dashed line (ap) represents the first
nonparabolic approximation to 3LM, the dashed-dotted line (IK) illustrates the results of Ref. [3] also based on
3LM (see text).}} \label{fig2th}
\end{figure}
In Fig. 2 we plot the spin $g^*$-value of the conduction electrons in bulk GaAs as a function of $d$ (or
$k_{z0}$). The solid line indicates the exact values of $g^*$ within 3LM, as obtained from Eqs. (11) and (12)
for $B$ = 2 T. The free-electron and distant band contributions are kept only for the conduction band. The
dashed line indicates values obtained from Eq. (17), resulting from the expansion for weak nonparabolicity. As
can be expected, for bigger widths $d$ (or smaller values of $k_{z0}$) the approximated values approach the
exact ones. Finally, the dashed line marked with IK indicates values obtained from Eq. (9) of Ref. [3] with the
use of the same material parameters. It can be seen that Ref. [3] predicts a much stronger dependence of $g^*$
on $k_{z0}$ than both our curves and, in particular, our dashed curve which has been calculated in the same
nonparabolic approximation as that used in Ref. [3]. The reason of this discrepancy is that the formulas given
in Ref. [3] correspond to our Eq. (17), with our $2m_0/m^*_0$ replaced by $-2E_P(1/G_0+2/E_0)/3$ and our $g^*_0$
replaced by $-2E_P(1/G_0-1/E_0)/3$. In case of the mass the approximation used in Ref. [3] is not too bad since
it misses only the term of 1+C [cf. our Eq. (18)]. In case of the $g^*$-value, however, it is seriously
inadequate since it misses the additive term of +2+2C', which is essential for the band-edge value of $g^*_0$ in
GaAs, see Eq. (19). As a result, the second term in the square bracket of Eq. (17) appears not with the
coefficient $g^*_0$ = -0.44, as it should, but with the coefficient of about -2.24. This gives a much too strong
dependence of $g^*$ on $k_{z0}^2$, as illustrated in Fig. 2.

The dependence of $g^*$ on $d$ (or $k_{z0}$), as given by the exact solution of 3LM (curve EX in Fig. 2), does
not describe experimental data in GaAs/Ga$_{0.67}$Al$_{0.33}$As quantum wells. In particular, it does not
exhibit the change of sign to positive $g^*$-factors for small values of $d$. On the other hand, we know that
the inclusion of higher conduction $\Gamma^v_7$, $\Gamma^v_8$ bands in the {\Pp} description (see Fig. 1)
results in stronger conduction band's nonparabolicity and, in particular, it gives the electron $g^*$-values
that change sign to positive at higher electron energies [1]. Thus it is necessary to include the $\Gamma^c_7$,
$\Gamma^c_8$ levels in the refined description of $g^*$ in GaAs. This is equivalent to using the full matrix (8)
for the determination of electron energies. In view of the applications to quantum wells the $k_z$ terms must be
retained. We pursue this program below.

\subsection{Five-level model}

In this section we calculate the spin $g^*$-value of conduction electrons using the five-level {\Pp} model. This
model is adequate for the GaAs-type materials. Once the two upper conduction levels are included, the matrix
elements $P_1$, $Q$ and $\overline\Delta$ come into play, see Fig. 1. As a result, two qualitatively new
features appear. First, the appearance of the matrix element $Q$ makes the conduction band nonspherical. Second,
the matrix element $Q$ does not vanish if the crystal is characterized by the bulk inversion asymmetry (BIA).
This results in the spin splitting of conduction energies for a given direction of the wave vector \textbf{k}
(the Dresselhaus mechanism [5]). We develop a workable approach to the calculation of $g^*$-factor in quantum
wells based of 5LM using an iterative expansion in powers of the interband matrix elements $P_0$, $P_1$ and $Q$.
Clearly, it is necessary to check whether such an expansion is adequate. In other words, we should make sure
that the expansion procedure gives the $g^*$-values comparable with those obtained from the complete matrix (8).
To this end we first solve the eigenenergy problem given by the {\Pp} Hamiltonian (8) by an alternative method,
first used by Evtuhov for the nonspherical valence bands of Ge [14]. As already mentioned, for spherical energy
bands the solutions of the eigenenergy problem in the presence of a magnetic field are given in terms of simple
harmonic oscillator functions. For not too strong deviations from band's sphericity we look for the envelope
functions in the form of sums of harmonic oscillator functions with unknown coefficients
\begin{equation}
f_l(\textbf{r})= exp(ik_zz)\sum_{m=0}c^l_m|m>\;\;.
\end{equation}
Since our Hamiltonian (8) contains raising and lowering operators for the harmonic oscillators, it is possible
to perform the prescribed operations. Then we multiply the obtained eigenenergy equations on the left by
consecutive harmonic oscillator functions and integrate the scalar products using the ortonormality relations.
This results in equations for the coefficients and the condition for nontrivial solutions gives the
eigenenergies. In other words, the Evtuhov method is a standard way of transforming a differential eigenvalue
problem into an algebraic problem by taking for the complete set of functions the harmonic oscillator functions.

Not going into details of the procedure we will give a general scheme for the resulting eigenenergy matrix. The
elementary irreducible block is a 7$\times$7 matrix, one containing the spin-up state and another the spin-down
state. If $Q$ and $k_z$ terms are neglected, the big matrix factorizes into the 7$\times$7 matrices for the
spin-up and spin-down energies for consecutive Landau levels. The nondiagonal $Q$ and $k_z$ terms couple the
elementary 7$\times$7 matrices. We find that in order to obtain precise energies for the lowest Landau levels we
need to go to 70$\times$70 matrices. The schematic form of such a matrix is given in Eq. (21).
\begin{widetext}
\begin{equation}
\left[ \begin{array}{cccccccccc}
N-1, -&Q\sqrt{B}&0&Qk_z&0&0&0&Pk_z&0&0 \\
&N, +&Q\sqrt{B}&Pk_z&Qk_z&0&0&Qk_z&0&0 \\
&&N+3, -&Qk_z&Pk_z&Qk_z&Q\sqrt{B}&0&0&0 \\
&&&N+1, -&Q\sqrt{B}&0&0&Q\sqrt{B}&0&0 \\
&&&&N+2, +&Q\sqrt{B}&Qk_z&0&0&0  \\
&&&&&N+5, -&Pk_z&0&Qk_z&Q\sqrt{B} \\
&&&&&&N+4, +&0&Q\sqrt{B}&Qkz  \\
&&&&&&& N-2, +&0& \\
&&&&&&&& N+7, -&Pk_z \\
&&&&&&&&&N+6, +
\end {array}
 \right]
\end{equation}
\end{widetext}
We use a symbolic notation, for example $N, +$ stands for the elementary 7$\times$7 matrix containing the
conduction Landau level $n$ with spin up. The symbols $Pk_z$ and $Qk_z$ stand for 7$\times$7 matrices containing
$P_0k_z$, $P_1k_z$ and $Qk_z$ terms, respectively, while $Q\sqrt{B}$ stands for 7$\times$7 matrices containing
$Q\sqrt{B}$ terms.

To calculate the $g^*$-value we will be interested in the lowest conduction levels $0^{\pm}$. For $N$=0, matrix
(21) gives the energy of the conduction level $0^+$ (second row and column) perturbed by interactions with other
levels. For $N$=1 we obtain the energy of the $0^-$ level (first raw and column). Fixing the values of $B$ and
$k_z$ we obtain from matrix (21) the energies which we consider to be "exact", so we use them as standards in
the estimation of following expansions.

\begin{table}
\caption{Band parameters of Ga$_{1-x}$Al$_{x}$As alloys for different chemical compositions $x$, as used in the
calculations, see [13, 15-18]. Energies E$_0$, G$_0$, E$_1$, G$_1$ are defined in Fig. 1, C and C' are far-band
contributions to the band-edge values of $m_0/m^*_0$ and $g^*_0$, respectively, see Eqs. (18) and (19). The
interband matrix elements of momentum and of the spin-orbit interaction are taken to be independent of $x$:
$E_{P_0}$=27.865 eV, $E_{P_1}$=2.361 eV, $E_{Q}$=15.563 eV, $\overline{\Delta}$=-0.061 eV.}
\begin{ruledtabular}
\begin{tabular}{cccc}
&GaAs&Ga$_{0.74}$Al$_{0.26}$As&Ga$_{0.67}$Al$_{0.33}$As\\
\hline
 G$_1$(eV) & 3.140 & 3.395 & 3.463 \\
 E$_1$(eV) & 2.969 & 3.221 & 3.289 \\
 E$_0$(eV) & -1.519 & -1.888 & -1.992 \\
G$_0$(eV)&-1.860&-2.199&-2.297\\
C&-2.297&-1.904&-2.248\\
C'&-0.025&-0.043&-0.056\\
V$_B$(eV)&---&0.208&0.264\\
m$^*_0$/m$_0$&0.066&0.0803&0.0875\\
g$^*_0$&-0.44&+0.40&+0.54\\
\end{tabular}
\end{ruledtabular}
\end{table}

Next we turn back to the set of 14 equations for the envelope functions $f_l$, as given by the initial
differential matrix (8). We find the envelope functions by substitution using an iteration procedure. In the
first step we put $Q$=0 and express twelve envelope functions by $f_4$ and $f_{11}$, corresponding to the
spin-up and spin-down $\Gamma^c_6$ conduction states. Next the $Q$ terms are restored and the zero-order
functions are put back to the complete equations. The twelve first-order $f_l$ functions are expressed by $f_4$
and $f_{11}$. In this approximation, linear in $Q$ terms, the BIA (Dresselhaus) spin splitting is included.
Finally, we return again to the initial set and determine the twelve functions $f_l$ by $f_4$ and $f_{11}$
including the $Q^2$ terms. The resulting equations for $f_4$ and $f_{11}$ represent the effective Hamiltonian.
It is obtained in the form of a 2$\times$2 differential matrix since the bulk inversion asymmetry mixes the two
spin states.

We are now in a position to calculate electron energies and the resulting conduction $g^*$-factor for given
values of $B$ and $k_z$ (still for the bulk). In the obtained formulas for energies there appear $k_z$, $k_z^2$
and $k_z^4$ terms. We checked that the $k_z$ and $k_z^4$ terms have negligible influence on the final energies.
\begin{figure}
\includegraphics[scale=0.55,angle=0]{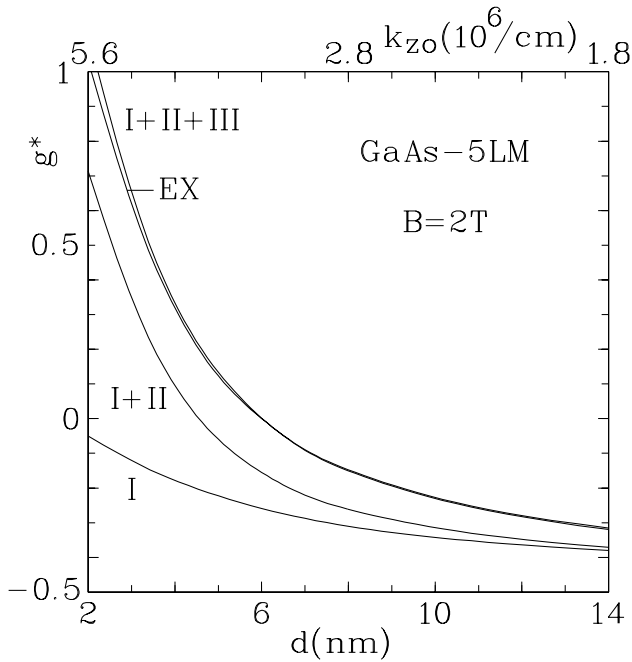}
\caption{\label{fig:epsart}{Theoretical spin $g^*$-factor of conducting electrons in bulk GaAs \emph{versus} the
wave vector $k_{z0}$, or the corresponding well width $d$ (see text), as calculated from the five-level model. I
- results for Q = 0 (first iteration step), I+II - bulk inversion asymmetry included (linear Q terms, second
iteration step), I+II+III - quadratic Q terms included (third iteration step). The line marked EX shows "exact"
results calculated numerically for the bulk by the Evtuhov method.}} \label{fig3th}
\end{figure}

In Fig. 3 we plot the calculated $g^*$-values as functions of $d$ (or $k_{z0}$) for the three consecutive
approximations. To determine $k_{z0}$ for a given width $d$ we use the same procedure as described before (cf.
Fig. 2). Our three approximations for the $g^*$-value correspond to the three iteration steps. Curve I
corresponds to $Q$=0, the corresponding formulas contain $P_0^2$ and $P_1^2$ type of terms. Curve I+II includes
the previous approximation plus the terms $P_0P_1Q$, $\overline{\Delta}P_0^2Q$, $\overline{\Delta}P_1^2Q$,
accounting for the spin splitting due to BIA. Curve I+II+III includes the two previous approximations plus the
terms $P_0^2Q^2$, $P_1^2Q^2$, $\overline{\Delta}P_0P_1Q^2$ (the last terms are very small and may be neglected).
In the same Fig. 3 we plot the $g^*$ value obtained from the "exact" numerical procedure based on matrix (21).

A discussion of the results shown in Fig. 3 is in order. First, curve I gives the results very similar to those
marked by EX in Fig. 2. Although the calculation I includes the higher conduction levels, as compared to 3LM
used for EX, it turns out that the results of 3LM and 5LM with $Q$=0 are very similar if one adjusts $C$ and
$C'$ constants to have the same band-edge values of $m^*_0$ and $g^*_0$. Second, sizable difference between the
calculations I and I+II may appear unexpected. Compared to I, the calculation I+II includes the Dresselhaus spin
splitting due to BIA, which is usually rather small. Indeed, for relatively large $d$ (small $k_{z0}$) the
corrections due to BIA are small. It is at small widths $d$ (large $k_{z0}$) that they become appreciable. One
should keep in mind that BIA works directly on the spin splitting, i.e. on the $g^*$ value. More generally, the
BIA spin splitting has somewhat particular properties since, while it is \emph{affected} by a magnetic field,
its primary origin is \emph{not} a magnetic field. Finally, it follows from Fig. 3 that the curve I+II+III
almost coincides with the "exact" result. This means that the third step in the iteration procedure gives enough
precision in the calculation of the electron $g^*$-value. We use this information in the following treatment of
the spin splitting in quantum wells.

\section{\label{sec:level3}$g^*$-FACTOR AND CYCLOTRON MASS IN QUANTUM WELLS\protect\\{}}

We apply the above results to calculation of the spin $g^*$-factor for conduction electrons in QWs grown along
the \textbf{z} direction. We use 5LM and consider both the growth direction and a magnetic field parallel to the
[001] crystal axis. Matrix (8) must now be completed by the potential $V(z)$ in all diagonal terms. The
calculation is carried out by iterating solutions of Eq. (8), as described above for the bulk. The difference is
that now we keep the operator $\hat{p} = (\hbar/i)\partial/\partial z$ in the differential form since the
envelope functions in QWs are unknown and they must be determined by solving the eigenenergy problem. As a
consequence, it is necessary in the iteration procedure to observe the order of $z$-dependent terms. In our
treatment we keep the free-electron and far-band contributions only in the conduction band. While a more general
treatment is possible, it makes the final formulas very lengthy while the precision is not markedly higher.

After a considerable manipulation the effective Hamiltonian for the spin-up and spin-down electron states in the
conduction band is obtained in the form
\begin{equation}
\hat{H}=\left[ \begin{array}{cc}
\hat{A}^+&\hat{K}\\
\hat{K}^\dagger&\hat{A}^-
\end {array}
 \right]\;\;,
\end{equation}
where
\begin{widetext}
$$
\hat{A}^+=V(z)-\frac{\hbar^2}{2}\frac{\partial}{\partial z} \frac{1}{m^*_I( {\cal E},
z)}\frac{\partial}{\partial z}+\frac{P^2_x+P^2_y}{2 m^*_I( {\cal E}, z)}+\frac{\mu_BB}{2}g^*_I( {\cal E}, z)
$$
$$
+\left(\frac{\hbar^2}{2m_0}\right)^2\frac{E_Q}{3\hbar^2} \left\{\frac{\partial}{\partial
z}\left[E_{P_0}(K_1P_-P_++K_2P_+P_-)+E_{P_1}(K_3P_-P_++K_4P_+P_-)\right]\frac{\partial}{\partial z}+ \right.
$$
\begin{equation}
\left. -\frac{2}{\hbar^2}\left[E_{P_0}(K_5P_-P_+P_-P_++K_6P_+^2P_-^2+K_7P_-^2P_+^2)+
E_{P_1}(K_8P_-P_+P_-P_++K_9P_+^2P_-^2+K_{10}P_-^2P_+^2)\right]\right\}\;\;,
\end{equation}
$$
\hat{A}^-=V(z)-\frac{\hbar^2}{2}\frac{\partial}{\partial z} \frac{1}{m^*_I( {\cal E},
z)}\frac{\partial}{\partial z}+\frac{P^2_x+P^2_y}{2 m^*_I( {\cal E}, z)}-\frac{\mu_BB}{2}g^*_I( {\cal E}, z)
$$
$$
+\left(\frac{\hbar^2}{2m_0}\right)^2\frac{E_Q}{3\hbar^2} \left\{\frac{\partial}{\partial
z}\left[E_{P_0}(K_1P_+P_-+K_2P_-P_+)+E_{P_1}(K_3P_+P_-+K_4P_-P_+)\right]\frac{\partial}{\partial z} \right.+$$
\begin{equation}
\left.-\frac{2}{\hbar^2}\left[E_{P_0}(K_5P_+P_-P_+P_-+K_6P_-^2P_+^2+K_7P_+^2P_-^2)+
E_{P_1}(K_8P_+P_-P_+P_-+K_9P_-^2P_+^2+K_{10}P_+^2P_-^2)\right]\right\}\;\;.
\end{equation}
The effective mass is
\begin{equation}
\frac{m_0}{m^*_I({\cal E}, z)}=1+C-\frac{1}{3}\left[E_{P_0}\left(\frac{2}{\tilde E_0}+ \frac{1}{\tilde
G_0}\right)+E_{P_1}\left(\frac{2}{\tilde G_1}+ \frac{1}{\tilde E_1}\right)
+\frac{4{\overline{\Delta}}\sqrt{E_{P_0}E_{P_1}}}{3}\left(\frac{1}{{\tilde E_1}{\tilde G_0}}- \frac{1}{{\tilde
E_0}{\tilde G_1}}\right)\right] \;\;,
\end{equation}
and the $g^*$-factor is
\begin{equation}
 g^*_I({\cal E}, z)=2+2C'+\frac{2}{3}\left[E_{P_0}\left(\frac{1}{\tilde E_0}- \frac{1}{\tilde
G_0}\right)+E_{P_1}\left(\frac{1}{\tilde G_1}- \frac{1}{\tilde
E_1}\right)-\frac{2{\overline{\Delta}}\sqrt{E_{P_0}E_{P_1}}}{3}\left(\frac{2}{{\tilde E_1}{\tilde G_0}}+
\frac{1}{{\tilde E_0}{\tilde G_1}}\right)\right] \;\;,
\end{equation}
\end{widetext}
where
\begin{equation}
{\tilde E_i} = E_i - {\cal{E}}+V(z)\;\;,
\end{equation}
\begin{equation}
{\tilde G_i} = G_i - {\cal{E}}+V(z)\;\;.
\end{equation}
We use the notation $E_{P_0}=2\hbar^2P^2_0/m_0$, $E_{P_1}=2\hbar^2P^2_1/m_0$ and $E_Q=2\hbar^2Q^2/m_0$. Further
$$
K_1=\frac{1}{3{\tilde G_1}}\left(\frac{26}{{\tilde G_0}{\tilde E_0}}+\frac{5}{{\tilde G_0}^2}+\frac{5}{{\tilde
E_0}^2}\right)\;\;,
$$
$$
K_2=\frac{1}{{\tilde E_0}^2}\left(\frac{8}{{\tilde E_1}}+\frac{1}{{\tilde G_1}}\right)+\frac{1}{{\tilde
G_1}{\tilde G_0}}\left(\frac{2}{{\tilde E_0}}+\frac{1}{{\tilde G_0}}\right)\;\;,
$$
$$
K_3=\frac{1}{3{\tilde E_0}}\left(\frac{26}{{\tilde G_1}{\tilde E_1}}+\frac{5}{{\tilde G_1}^2}+\frac{5}{{\tilde
E_1}^2}\right)\;\;,
$$
$$
K_4=\frac{1}{{\tilde G_1}^2}\left(\frac{8}{{\tilde G_0}}+\frac{1}{{\tilde E_0}}\right)+\frac{1}{{\tilde
E_1}{\tilde E_0}}\left(\frac{2}{{\tilde G_1}}+\frac{1}{{\tilde E_1}}\right)\;\;,
$$
$$
K_5=\frac{1}{3{\tilde G_1}}\left(\frac{1}{{\tilde E_0}}-\frac{1}{{\tilde G_0}}\right)^2\;\;,
$$
$$
K_6=\frac{1}{2{\tilde E_0}^2}\left(\frac{1}{{\tilde E_1}}+\frac{2}{{\tilde G_1}}\right)\;\;,
$$
$$
K_7=\frac{1}{2}\left(\frac{2}{{\tilde G_1}{\tilde G_0}^2}+\frac{1}{{\tilde E_1}{\tilde E_0}^2}\right)\;\;,
$$
$$
K_8=\frac{1}{3{\tilde E_0}}\left(\frac{1}{{\tilde E_1}}-\frac{1}{{\tilde G_1}}\right)^2\;\;,
$$
$$
K_9=\frac{1}{2{\tilde G_1}^2}\left(\frac{1}{{\tilde G_0}}+\frac{2}{{\tilde E_0}}\right)\;\;,
$$
\begin{equation}
K_{10}=\frac{1}{2}\left(\frac{2}{{\tilde E_0}{\tilde E_1}^2}+\frac{1}{{\tilde G_0}{\tilde G_1}^2}\right)\;\;.
\end{equation}

We emphasize that the mass $m^*_I$ and the g-value $g^*_I$ defined in Eqs. (25) and (26) do not represent the
final cyclotron mass and g-value in a quantum well but only the first iterative approximations to these
quantities, as obtained from matrix (8) by putting Q = 0.

The nondiagonal component in Eq. (22), related to the bulk inversion asymmetry, is
\begin{equation}
{\hat K}={\hat B}_1 + {\hat B}_2\;\;,
\end{equation}
where
\begin{equation}
{\hat B}_1=\frac{-\sqrt{2}}{\hbar}\gamma( {\cal E}, z)\left[P_+\frac{\partial^2}{\partial z^2}
+\frac{1}{4\hbar^2}(P_-P_+^2+P_+^2P_-) \right]\;\;,
\end{equation}

\begin{equation}
{\hat B}_2=\frac{1}{{\sqrt 2}\hbar^3}\gamma( {\cal E}, z)P_-^3\;\;,
\end{equation}
in which
\[
\gamma({\cal E}, z)=\frac{4Q}{3} \left\{P_0P_1\left(\frac{1}{\tilde G_0 \tilde G_1} -\frac{1}{\tilde E_0 \tilde
E_1}\right)+ \right.
\]
\begin{equation}
\left. -\frac{\overline{\Delta}}{3}\left[ \frac{P^2_0}{\tilde E_0 \tilde G_0}\left(\frac{2}{\tilde E_1}+
\frac{1}{\tilde G_1}\right)- \frac{P^2_1}{\tilde E_1 \tilde G_1}\left( \frac{2}{\tilde G_0}+\frac{1}{\tilde
E_0}\right)\right]\right\}\;\;.
\end{equation}

The diagonal components in the Hamiltonian (22) are composed of the terms proportional to $P_0^2$ and $P_1^2$,
resulting from the first iteration step (marked I in the previous sections). They contain also the terms
proportional to $P_0^2Q^2$ and $P_1^2Q^2$, resulting from the third iteration step (marked III). The diagonal
terms neither raise nor lower the harmonic oscillator functions. The nondiagonal terms ${\hat B}_1$ and ${\hat
B}_2$ are proportional to $Q$ and they come from the second iteration step (marked II). They result from the
bulk inversion asymmetry of the crystal. The operator ${\hat B}_1$ raises the harmonic oscillator function $|n>$
to $|n+1>$, whereas ${\hat B}_2$ lowers it from $|n>$ to $|n-3>$. In consequence, solving the eigenenergy
problem imposed by Eq. (22) requires again the Evtuhov procedure. However, now we have to look for solutions in
the form
\begin{figure}
\includegraphics[scale=0.55,angle=0]{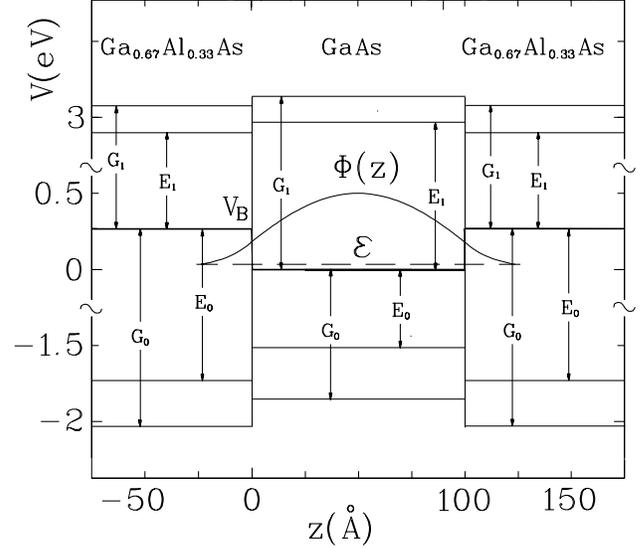}
\caption{\label{fig:epsart}{Band structure of a rectangular GaAs/Ga$_{0.67}$Al$_{0.33}$As quantum well within
5LM along the growth direction $z$. The calculated energy and wavefunction of the ground subband is indicated.}}
\label{fig4th}
\end{figure}

\begin{equation}
F_l= \sum_{m=0}^\infty c^l_m|m>\chi_m(z)\;\;,
\end{equation}
where $l$=1, 2, the functions $|m> = A_m exp(ik_xx)\Phi_m(y)$ are the same as for the bulk, but $\chi_m(z)$ are
as yet unknown envelope functions describing the motion along the growth direction $z$. When applying the
functions (34) we will limit them to the minimal coupling scheme. It can be seen from Eq. (22) that the spin-up
state $|n, +>$ described by ${\hat A}^+$ interacts via the offdiagonal elements ${\hat B}_1$ and ${\hat B}_2$
with the spin-down states. In view of the above considerations it is clear that $|n, +>$ interacts via ${\hat
B}_2$ with $|n+3, ->$ and via ${\hat B}_1$ with $|n-1, ->$. After performing the operations on the harmonic
oscillator functions we obtain the following eigenenergy problem in the minimal coupling scheme
\begin{equation}
 \left( \begin{array}{ccc}
 \hat{A}^+_n-{\cal E}&\hat{B}_{2, n}&\hat{B}_{1, n} \\
\hat{B}^\dagger_{2, n}&\hat{A}^-_{n+3}-{\cal E}&0 \\
\hat{B}^\dagger_{1, n}&0&\hat{A}^-_{n-1}- {\cal E}
 \end{array} \right)
\left( \begin{array}{c} \chi_n(z) \\ \chi_{n+3}(z)\\ \chi_{n-1}(z) \end{array} \right) =0\;\;.
\end{equation}

This set of coupled differential equations gives three energies, of which we are interested in ${\cal E}(n, +)$.
A similar reasoning leads to the following set of differential equations containing the $|n, ->$ state
\begin{equation}
 \left( \begin{array}{ccc}
 \hat{A}^+_{n-3}-{\cal E}&0&\hat{B}_{2, n} \\
0&\hat{A}^+_{n+1}-{\cal E}&\hat{B}_{1, n} \\
\hat{B}^\dagger_{2, n}&\hat{B}^\dagger_{1, n}&\hat{A}^-_n- {\cal E}
 \end{array} \right)
\left( \begin{array}{c} \chi_{n-3}(z) \\ \chi_{n+1}(z)\\ \chi_n(z) \end{array} \right) =0,
\end{equation}
Here we are interested in ${\cal E}(n, -)$.

In the above matrices we use the following notation
\begin{widetext}
$$
\hat{A}^+_n=V(z)-\frac{\hbar^2}{2}\frac{\partial}{\partial z} \frac{1}{m^*_I({\cal E},
z)}\frac{\partial}{\partial z}+\frac{\hbar e B}{m^*_I({\cal E}, z)}(n+1/2)+\frac{\mu_BB}{2}g^*_I( {\cal E}, z) +
$$
$$
+E_Q\frac{\hbar\omega_c^0}{6}\left\{\frac{\hbar^2}{2m_0}\frac{\partial}{\partial z}\left[E_{P_0}\left(
(n+1)K_1+nK_2\right)+E_{P_1}\left( (n+1)K_3+nK_4\right)\right]\frac{\partial}{\partial z}+\right.
$$
\begin{equation}
\left.
-\hbar\omega_c^0\left[E_{P_0}\left((n+1)^2K_5+n(n-1)K_6+(n+1)(n+2)K_7\right)+E_{P_1}\left((n+1)^2K_8+n(n-1)K_9+
(n+1)(n+2)K_{10}\right)\right]\right\}
\end{equation}

$$
\hat{A}^-_n=V(z)-\frac{\hbar^2}{2}\frac{\partial}{\partial z} \frac{1}{m^*_I({\cal E},
z)}\frac{\partial}{\partial z}+\frac{\hbar e B}{m^*_I({\cal E}, z)}(n+1/2)-\frac{\mu_BB}{2}g^*_I( {\cal E}, z) +
$$
$$
+E_Q\frac{\hbar\omega_c^0}{6}\left\{\frac{\hbar^2}{2m_0}\frac{\partial}{\partial z}\left[E_{P_0}\left(
nK_1+(n+1)K_2\right)+E_{P_1}\left(nK_3+(n+1)K_4\right)\right]\frac{\partial}{\partial z}+\right.
$$
\begin{equation}
\left.
-\hbar\omega_c^0\left[E_{P_0}\left(n^2K_5+(n+1)(n+2)K_6+n(n-1)K_7\right)+E_{P_1}\left(n^2K_8+(n+1)(n+2)K_9+
n(n-1)K_{10}\right)\right]\right\}
\end{equation}
\end{widetext}

\begin{equation}
{\hat B}_{1, n}=\gamma({\cal E}, z)\frac{\sqrt{2}}{L}\left(\sqrt{n+1}\frac{\partial^2}{\partial z^2}
+\frac{(n+1)\sqrt{n+1}}{2L^2}\right)
\end{equation}

\begin{equation}
{\hat B}_{2, n}=-\frac{\gamma( {\cal E}, z)}{\sqrt{2}L^3}\sqrt{(n+3)(n+2)(n+1)}
\end{equation}
in which $K_1,...K_{10}$ are again given by Eq. (29).

If one is interested in the lowest spin states $|0, \pm>$, sets (35) and (36) do not contain the $n-1$ and $n-3$
components and the eigenenergy problems reduce to the following two sets of coupled differential equations. For
the $|0, +>$ state
\begin{equation}
 \left( \begin{array}{cc}
 \hat{A}^+_0-{\cal E}&\hat{B}_{2, n} \\
\hat{B}^\dagger_{2, n}&\hat{A}^-_3-{\cal E}
 \end{array} \right)
\left( \begin{array}{c} \chi_0(z) \\ \chi_3(z)\end{array} \right) =0\;\;,
\end{equation}
while for the $|0, ->$ state
\begin{equation}
 \left( \begin{array}{cc}
\hat{A}^+_1-{\cal E}&\hat{B}_{1, n} \\
\hat{B}^\dagger_{1, n}&\hat{A}^-_0- {\cal E}
 \end{array} \right)
\left( \begin{array}{c}\chi_1(z)\\ \chi_0(z) \end{array} \right) =0\;\;.
\end{equation}
In order to determine the cyclotron mass for the lowest levels we need to know in addition the energies of $|1,
+>$ and $|1, ->$ states. As follows from Eq. (36), for the $|1, ->$ state the corresponding differential set
still contains two equations, while for the $|1, +>$ state we have in principle three coupled equations, see Eq.
(35). However, we found that the addition of the third state $|4, ->$ (see Eq. (35)) does not really change the
energy of the $|1, +>$ state, so that two equations suffice to determine its energy.

Once the electron is in a QW of a finite height, its total $g^*$-factor is given not only by its properties in
the well but also by those in the barriers. The situation for a rectangular GaAs/Ga$_{0.67}$Al$_{0.33}$As QW is
shown in Fig. 4. Since the electron wave function penetrates into the barriers in which $g^*$ is different from
those in the well, we should average over the two regions. Seemingly, one should simply use the band edge values
of $g^*$ in GaAs and Ga$_{0.67}$Al$_{0.33}$As and take an appropriate average. In fact, this was done in Ref. 3.
We argue, however, that this is incorrect. The reason is that, as follows from Fig. 4, the electron energy is
much lower than the barrier's band edge $V_B$. This effect is automatically included in our formalism if we use
for both the well and the barriers the appropriate functions ${\tilde E_i}$ and ${\tilde G_i}$, as given by Eqs.
(27) and (28), respectively. When considering the two regions we have to change not only the appropriate values
of energy gaps and spin-orbit energies but also the potential $V(z)$. At the flat bottom of the well shown in
Fig. 4, there is $V(z)$=0. On the other hand, in the barrier regions there is $V(z)=V_B$, where $V_B$ is the
offset value for the conduction band between the two materials. Clearly, the energy ${\cal E}$ is the same in
the two regions. Only if $V_B$ were zero would we deal with "straight" values of $g^*$ in both materials.

It is instructive to consider a numerical example. Taking for Ga$_{0.67}$Al$_{0.33}$As the parameters given in
Table 1 we obtain the band edge value of $g^*_0$ = 0.5378, in agreement with the experimental findings [13]. On
the other hand, when we compute the value of $g^*$ in the barrier using the same parameters but accounting for
$V_B$ = 0.264meV and taking to a good approximation ${\cal E} \approx$ 0, we obtain $g^*$ = 0.1638, which is a
considerably lower value.

In fact, in our formalism we do not employ the averaging procedure. When calculating the total electron $g^*$
value we use in the expressions for $g^*_I({\cal E}, z)$, $m^*_I({\cal E}, z)$ and $K_i$, as given by Eqs (37),
(38) and (29), the actual functions ${\tilde E_i}$ and ${\tilde G_i}$ for a given region. This includes $V(z)$
and the local values of band parameters. Thus the actual values of $g^*_I$, $m^*_I$ and $K_i$ enter both the
differential equations defined by Eq. (22), as well as the boundary conditions at the interfaces. The problem of
the boundary conditions was intensely discussed in the early days of heterostructure investigations, see [4]. It
will suffice to say here that our Eqs. (37) and (38) are written in the hermitian form, so that the boundary
conditions are obtained in the form of continuity of the wave function and of electric current across the
interfaces. The eigenenergy calculation is done for each spin separately so that, as follows from Eq. (22), also
the boundary conditions depend on the spin. Since the effective equations for the energies contain only $p^2_z$
terms, the boundary conditions are not more complicated than those for a $z$-dependent effective mass. Once
${\cal E}^+_0$ and ${\cal E}^-_0$ are computed, the $g^*$-value is determined using the definition (15).

All the above remarks apply also to the computation of the cyclotron mass. The mass value in barriers also
contains the offset $V_B$. In the above numerical example for Ga$_{0.67}$Al$_{0.33}$As, instead of the band-edge
value $m^*_0$ = 0.0875$m_0$
 we deal with the value $m^*$ = 0.0748$m_0$ in the barrier. We emphasize that this feature is
of importance even in the simplest calculation of the energy and of the effective mass in a QW made of two
materials. The above problem is recognized in the work of Bastard et al [19, 20] as well as in our earlier
papers [21, 22].

\section{\label{sec:level2}RESULTS AND DISCUSSION}

Figure 5 shows our calculation of the electron spin $g^*$-factor in GaAs/Ga$_{0.67}$Al$_{0.33}$As rectangular
QWs together with available experimental data. This is the main result of our work. It can be seen that the
complete five - level {\Pp} model gives an excellent description of the $g^*$-value as a function of the well
width between 3nm and 12nm. In particular, the theory reproduces the experimental value of $g^*$ = 0 for the
well width of 6 nm, as observed by Le Jeune et al [23]. In the discrepancy of the experimental points around d
$\approx$ 4.5 nm, priority should be given to Ref. [23] (the full circle) since the Raman data of Ref. [24] (the
empty circle) do not determine the conduction $g^*$-value directly. Our calculation is carried out for the
growth direction $\textbf{z} ||[001]$ with an external magnetic field parallel to this direction. In order to
reach the good description of experiments we had to use the full five-level model of the band structure, as
discussed above. The theory agrees perfectly well with the experimental $g^*$-value even for $d \approx$ 30
$\rm{\AA}$. One could find it surprising that the {\Pp} theory works so well on the scale of few interatomic
distances in GaAs, but this result is in agreement with a more general experience that the {\Pp} theory works
better than it should.
\begin{figure}
\includegraphics[scale=0.55,angle=0]{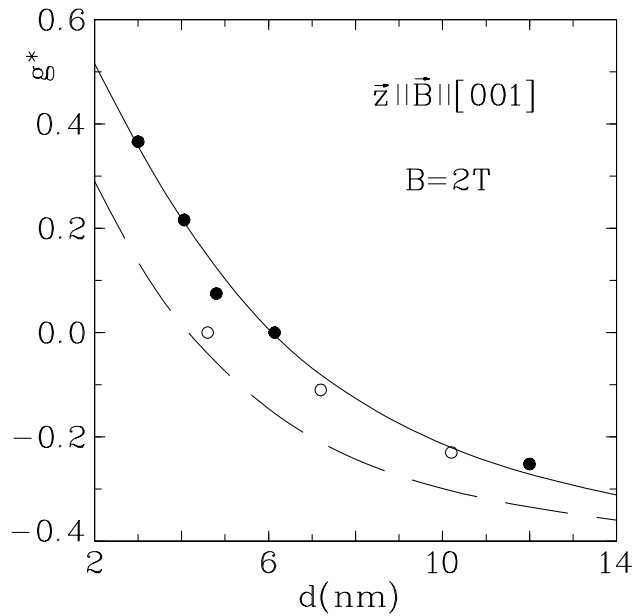}
\caption{\label{fig:epsart}{Spin $g^*$-factor of conduction electrons in rectangular
GaAs/Ga$_{0.67}$Al$_{0.33}$As quantum wells \emph{versus} well width $d$. The solid line - theory based on 5LM,
the dashed line - theory based on 3LM, full circles - experimental data of Le Jeune et al [23], empty circles -
experimental data of Sapega et al [24].}} \label{fig5th}
\end{figure}
It can be seen that the total $g^*$-value for narrow wells is smaller than the value for pure GaAs, as shown in
Fig. 5. This is due to the above mentioned effect of the offset $V_B$, which lowers the effective $g^*$ in the
barriers, as compared to the band-edge $g^*$-value in Ga$_{0.67}$Al$_{0.33}$As.

The dashed line in Fig. 5 shows the calculation of $g^*$-factor, in which we used the three-level {\Pp} model.
This model does not involve the $Q$ matrix element, so that it includes neither BIA splitting nor the $Q^2$
terms. It can be seen that 3LM does not properly describe the experimental data in GaAs/Ga$_{0.67}$Al$_{0.33}$As
quantum wells. This result agrees with the general conclusion that the three-level model does not describe
correctly the conduction band in GaAs - type materials, see [1, 2]. The values of $g^*$ shown in Fig. 5 were
computed for the magnetic field intensity $B$ = 2 T. It is natural that in a theory for the nonparabolic energy
band, when the energies are nonlinear functions of a magnetic field, the $g^*$-value may depend somewhat on $B$.
We checked this dependence and found that diminishing the field until $B \approx$ 0.25 T the $g^*$-value varies
only weakly with $B$. However, as $B$ tends to zero the $g^*$-value ceases to be a useful quantity because of
the spin splitting due to BIA. It is then more practical to use the energy splitting for the two spins.

As to the previous description of $g^*$-value in QW based on 3LM [3], it is incorrect on two accounts. First, as
already mentioned in the discussion of Fig. 2, it predicts a much too strong increase of $g^*$ in GaAs with
decreasing well width $d$. Second, when averaging the spin $g^*$-factors in GaAs well and
Ga$_{0.67}$Al$_{0.33}$As barriers, it takes the band-edge values in both materials, neglecting the influence of
the band offset on the $g^*$-value in barriers. Both above shortcomings lead to an increase of the total $g^*$.
We conclude that the agreement of the theory [3] with the experiment on GaAs/Ga$_{0.67}$Al$_{0.33}$As QW, as
quoted by Le Jeune et al [23], is fortuitous.

\begin{figure}
\includegraphics[scale=0.55,angle=0]{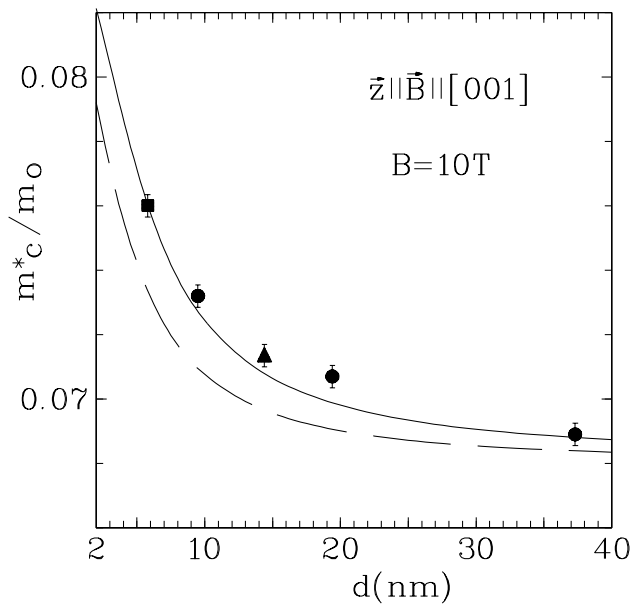}
\caption{\label{fig:epsart}{Cyclotron mass of conduction electrons in rectangular GaAs/Ga$_{0.74}$Al$_{0.26}$As}
quantum wells \emph{versus} well width $d$. The solid line - theory based on 5LM, the dashed line - theory,
based on 3LM, experimental data are after Huant et al [25] for samples with slightly different chemical
composition. Full circles x=0.26, full square x=0.25, full triangle x=0.23.} \label{fig6th}
\end{figure}

In Fig. 6 we plot the cyclotron mass of electrons in GaAs/Ga$_{0.74}$Al$_{0.26}$As rectangular QWs as a function
of the well width according to the three-level and five-level models. It can be seen that 3LM gives distinctly
lower masses, although we adjust the far-band contribution $C$ to get the same band-edge value of $m^*_0$ =
0.0665 $m_0$. The five-level model gives slightly different masses for spin-up and spin-down cyclotron
transitions.

Figure 6 shows the comparison of our theory with the cyclotron resonance data of Huant et al [25] obtained on
GaAs/Ga$_{1-x}$Al$_x$As QWs. For the GaAs/Ga$_{0.74}$Al$_{0.26}$As alloy we use the band parameters indicated in
Table 1. For the band edge in GaAs we take the perfectly acceptable value of $m^*_0$ = 0.0665 $m_0$, accounting
for the fact that the data of Ref. [25] were taken at the temperatures 20-30 K. Our theory describes the data
very well with the exception of the point at $d \approx$ 20 nm, but this point does not follow well the overall
mass behavior. The calculation shown in Fig. 6 was performed for $B$ = 10 T, which corresponds to the
experimental conditions of Ref. [25]. Comparing the data with the theory we conclude that the three-level model
can not correctly describe the experimental data.

The previous theoretical treatment of the effective masses in GaAs/Ga$_{1-x}$Al$_x$As QWs by Ekenberg [4], which
was carried out for $B$ = 0, makes a distinction between the masses parallel and perpendicular to the growth
direction. We do not introduce the parallel mass since it does not seem to have a clear physical meaning. Also,
in nonparabolic energy bands one can not separate the electron motion in different directions since an increase
of the mass due to motion in one direction will affect the motion in other directions. The theory of Ref. [4] is
based on an expansion of the energy in powers of momentum [26]. The coefficients in such an expansion contain
the band-edge energies in their denominators and the interband matrix elements of momentum in their numerators.
In Ref. [4] the potential $V(z)$ of the well is added later, so that this procedure can be qualified as
semiclassical. On the other hand, in our treatment we introduce the potential $V(z)$ and the energy $\cal E$
from the beginning and use the iteration procedure in solving the resulting 14 equations by substitution. This
procedure is exact for the matrix element $Q$ = 0 and includes $Q$ to second order. The resulting effective
equations have the coefficients which not only contain the band-edge energies, but also involve in their
denominators $V(z)$ and $\cal E$, cf. Eqs. (27), (28), (29). We showed above that these quantities are not
negligible in GaAs-type materials and they would be even more important in narrow-gap materials. The work [4]
does not include the bulk inversion asymmetry effects, which are not very important for the cyclotron mass but
become important for the spin $g^*$-value, see our Fig. 3. Our theoretical masses for
GaAs/Ga$_{0.7}$Al$_{0.3}$As QW (not shown) are somewhat higher then those presented in Fig. 4 of Ref. [4]. It is
not clear whether the differences result from the approximations made in Ref. [4] or from somewhat different
band parameters used in the two calculations.

The approach presented above can be used equally well for quantum wells described by an arbitrary potential
$V(z)$. This amounts to solving differential equations (22), (37), (38), in which $V(z)$ explicitly appears.
However, if a well (and a corresponding potential) does not possess the inversion symmetry, there appears a spin
splitting due to the Bychkov-Rashba mechanism [6]. Technically, this appears as a result of noncommutativity of
$\hat{p}_z$ with $V(z)$ when one solves the {\Pp} equations by substitution. The Bychkov-Rashba splitting has
been treated in many papers, see the review [27].

The present work confirms our previous description of the spin $g^*$-factor in parabolic GaAs/Ga$_{1-x}$Al$_x$As
quantum wells [28, 29]. We conclude that the five-level {\Pp} model adequately describes this important system.

\section{\label{sec:level2}SUMMARY}

We describe the spin and cyclotron energies of electrons in GaAs/Ga$_{1-x}$Al$_x$As quantum wells in the
presence of an external magnetic field parallel to the growth direction [001]. Our approach is based on the
five-level {\Pp} model of band structure for GaAs-type materials. Inadequacy of the previous theory used in the
literature for the spin $g^*$-factor in heterostructures is indicated. We solve 14 coupled differential
equations resulting from the {\Pp} formulation by three iteration steps. The sufficiency of this procedure is
tested on bulk GaAs energies for different values of $k^2_z$ corresponding to real quantum wells. Influence of
the bulk inversion asymmetry present in III-V compounds on the electron $g^*$-value is emphasized. Our theory
gives an excellent description of the existing experimental data on the spin $g^*$-factor and the cyclotron mass
of electrons in GaAs/Ga$_{1-x}$Al$_x$As rectangular quantum wells for different well widths.

\begin{acknowledgments} We are grateful to Professor A. Raymond for informative discussions. This work was supported in part by The Polish Ministry of Sciences, Grant No
PBZ-MIN-008/PO3/2003.
\end{acknowledgments}

\end{document}